# Cd(Zn)O on SiC: epsilon-near-zero modes and plasmon-phonon coupling

Maria Villanueva-Blanco (maria.villanueva@upm.es) [a], Javier Yeste (Javier.Yeste@uv.es) [b], Julia Ingles-Cerrillo (julia.ingles@upm.es) [a], Pablo Ibañez-Romero (p.iromero@upm.es) [a], Carmen Martínez-Tomas (Carmen.Martinez-Tomas@uv.es) [b], Vicente Muñoz-Sanjosé (Vicente.Munoz@uv.es) [b], Miguel Montes Bajo (miguel.montes@upm.es) [a] and Adrian Hierro (adrian.hierro@upm.es) [a]*

[a] ISOM, Universidad Politécnica de Madrid, 28040 Madrid, Spain

[b] Dept. Física Aplicada i Electromagnetisme, Universitat de València, 46100 Burjassot, Valencia, Spain

* To whom correspondence should be addressed.

## Abstract

Cd(Zn)O stands out as probably the best plasmonic material in the mid-IR, but it is usually grown on sapphire or other passive substrates. In this work we introduce SiC as a novel, highly polar, dopable substrate for Cd(Zn)O. The Cd(Zn)O/SiC system is analyzed as a function of the Zn concentration and thin film thickness, and the results are compared to those obtained in the Cd(Zn)O/sapphire system. XRD and reflectance measurements show that the alloy with 10 % Zn nominal concentration has the best crystalline and plasmonic quality, with optical losses as good as 13 % of the plasma frequency. The thin films show two surface polariton modes: a purely plasmonic symmetric mode at higher energies with negligible frequency dispersion and pinning at the plasma frequency for the thinnest films, characteristic of an ideal epsilon-near-zero mode; and a plasmonic-phononic hybridized antisymmetric mode at lower energies, which thanks to the large value of the higher frequency dielectric constant of SiC compared to sapphire, shows much lower frequency dispersion, indicative of the stronger epsilon-near-zero character. Hence, Cd(Zn)O/SiC offers a promising platform for the development of ENZ devices on an active substrate.

# 1. Introduction

Materials with a refractive index approaching zero have attracted significant interest in the last decade due to their unusual optical response like large field enhancement inside the material [1], slow-light [2] and nonlinearity [3]. In the case of non-magnetic materials with unity relative permeability, these conditions are achieved when the real portion of the permittivity approaches zero ($|Re\{\varepsilon\}| < 1$), and they are labeled epsilon-near-zero (ENZ) materials. ENZ materials have been used to control tunneling [4], switching [5] and to develop perfect absorbers [6].

Among the different spectral regions, the mid-infrared (mid-IR) range offers significant opportunities for ENZ phenomena due to its potential applications in chemical sensing [7,8], high-harmonic generation [9] and optical analysis [10]. The search for appropriate materials with resonances in the mid-IR has been exhaustive, focusing on doped semiconductors [11,12] and metamaterials [3]. Transparent conductive oxides (TCOs) have shown the best figures of merit in the mid-IR range [13,14], and among them, doped or alloyed CdO stands out with the lowest optical losses [15,16]. CdO has a tunable plasma frequency that can be changed with the alloying of other metals like Zn [17] and doping [18]. Thanks to its low-losses and controllable high carrier concentration [19], Cd(Zn)O shows a great potential for mid-IR applications.

The plasmonic properties of CdO and alloyed Cd(Zn)O thin films have been analyzed on a great variety of substrates like Si [18], MgO [20], GaAs [21], and sapphire [15,19]. On sapphire, CdO thin films reportedly grow with a high crystalline quality in most of the plane orientations [22]. Additionally, the optical phonons of sapphire and the plasmon of Cd(Zn)O combine to form a hybrid surface plasmon-phonon polariton, a highly propagative low-loss mode that could be used for molecular sensing and the improvement of mid-IR waveguides [23]. Despite the advantages sapphire offers, it is very insulating, which can be a serious limitation for the integration of CdO plasmonic structures on active devices. Hence, it would be interesting to expand the search towards other active substrates that can offer new opportunities in the development of micro- or optoelectronic devices with integrated CdO.

In this context, SiC is an interesting substrate that can cover these needs. It is a wide-gap polar semiconductor whose electrical characteristics can be manipulated by doping. It is present in a variety of commercial devices such as SiC MOSFETs [24], where it is the active material, or GaN HEMTs [25], where it is chosen due to its high thermal conductivity. In this work we report high quality alloyed Cd(Zn)O thin films grown on SiC substrates. We evaluate the plasmonic properties of the thin films according to the alloy content and thin film thickness and we achieve a sufficiently thin film to obtain a hybridized surface plasmon-phonon polariton mode and ENZ modes at the Cd(Zn)O plasma and near the SiC phonon frequencies.

## 2. Results and discussion

$Cd_{1-x}Zn_xO$ thin films were grown on c-plane 4H-SiC substrates using metal organic chemical vapor deposition (MOCVD). Two sets of samples were grown. The first set (Set 1) considers films with a thickness of around 400 nm and a variation in Zn nominal content between 0 and 20 %. The second set (Set 2) considers a fixed 10 % Zn content and a thin film thickness between 20 and 380 nm.

The Set 1 of samples were measured by X-ray diffraction and the corresponding rocking curves were used to evaluate the crystalline quality of the films. As seen in Figure 1a, The FWHM of the Cd(Zn)O(111) peak decreases when increasing the Zn content from 0 to 10 %, and increases again for the largest Zn content. We find the best crystal quality in the samples with a nominal 10 % Zn content. Figures 1b-c show scanning-electron microscopy (SEM) images of the sample with 10 % Zn content. As seen in the images, the deposited Cd(Zn)O assembles into a compact film with some rugosity. Atomic Force Microscopy (AFM) measures of the surface show a root mean square (RMS) roughness on the order of 9 nm, similar to the one reported on sapphire substrates [26].

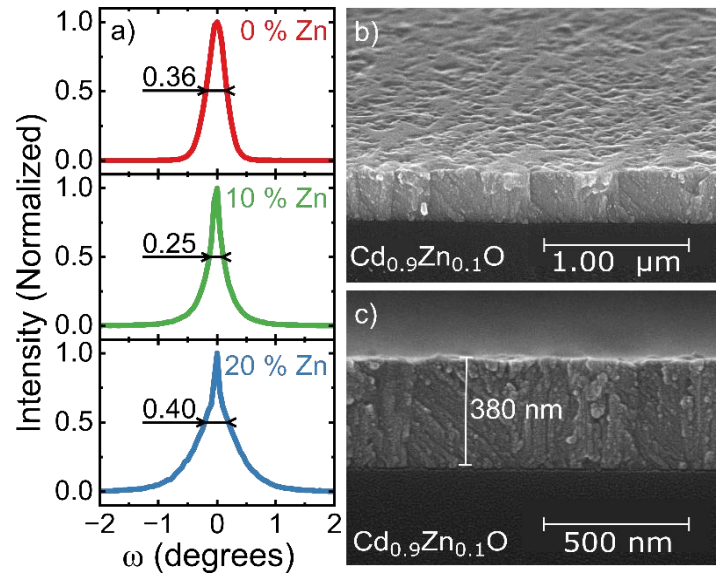

Figure 1: a) Rocking curves of Cd(111) peaks corresponding to different Zn contents. The black labels inside the plots indicate the FWHM. b-c) SEM images of the cross section of the sample with 10 % Zn content.

In order to extract the screened plasma frequency ($\omega_p$), damping ($\gamma_p$), and film thickness related to each sample, polarized reflectance spectroscopy was used, and the experimentally obtained reflectance spectra were fitted with a Transfer Matrix Method (TMM) simulation (see Methods). Figure 2a shows both the experimental and fitted spectra from which the thin film properties are extracted (Table 1).

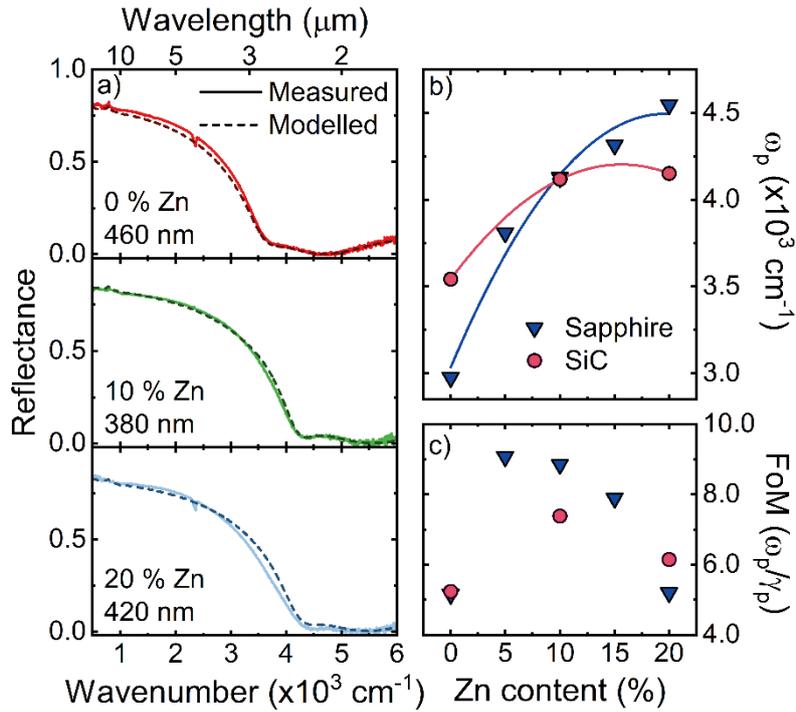

Figure 2: a) Reflectance of samples with varying Zn content. The solid line represents the experimental measurement taken under p-polarization at an incident angle of 45º; the dashed line represents the TMM modelled fit. The film thickness was obtained from the fit. b) Screened plasma frequency ($\omega_p$) as a function of Zn content for all the samples in Set 1 as a function of the Zn content in Cd(Zn)O thin films over SiC (red) or sapphire (blue) substrates. The solid lines are a guide to the eye. The data for Cd(Zn)O on sapphire was extracted from [17]. c) Figure of Merit (FoM) of samples with different Zn content in Cd(Zn)O thin films over SiC (red) or sapphire (blue) substrates. The data for the calculation of the FoM in the sapphire substrate was obtained from the PhD thesis by Julen Tamayo-Arriola [27].

|  | Cd(Zn)O thickness (nm) | Nominal % Zn | $\omega_p$ (cm$^{-1}$) | $\gamma_p$ (cm$^{-1}$) |
|---|---|---|---|---|
| Set 1 | 460 | 0 | 3543 | 677 |
| Set 1 | 380 | 10 | 4117 | 558 |
| Set 1 | 420 | 20 | 4151 | 675 |
| Set 2 | 20 | 10 | 4139 | 750 |
| Set 2 | 200 | 10 | 3968 | 512 |
| Set 2 | 380 | 10 | 4117 | 558 |

Table 1: Cd(Zn)O thin film properties grouped as a function of Zn content (Set 1), and thickness (Set 2). All the parameters in this table except the nominal Zn content were obtained by fitting a simulation obtained from the TMM method to the experimental reflectance.

Figure 2b shows $\omega_p$ for each of the measured samples (red circles). The results are compared to those of thin Cd(Zn)O films grown on a sapphire substrate (blue triangles), as reported by Tamayo-Arriola [17]. As shown in the figure, for both substrates $\omega_p$ stays in the 3000-4500 cm$^{-1}$ range (2.2-3.3 µm). However, the mean value for $\omega_p$ varies significantly more with Zn content on sapphire substrates than on SiC substrates. In both cases, $\omega_p$ saturates at higher Zn nominal concentrations. For the case of the unalloyed CdO with no Zn, $\omega_p$ is higher for the SiC substrate than for the sapphire substrate.

In order to determine the plasmonic quality of the thin films we use a figure of merit defined as $\text{FoM} = \omega_p/\gamma_p$. In Figure 2c, a comparison is made between the best FoM in samples with a SiC substrate (red circles), and with a sapphire substrate (blue triangles) [17,27]. In SiC the FoM ranges from 5.2 to 7.8, it reaches the best results in the 10 % Zn alloy, corresponding to a damping as low as 13 % of the plasma frequency, and decreases for higher concentrations, as does the FoM in sapphire. This result is consistent with the analysis of crystalline quality obtained in Figure 1: The better the crystalline quality of the sample, the better the figure of merit. The values are among the best of those obtained in the same frequency range in previous studies done in other semiconductors [28] .

Having established that the thin film with 10 % Zn content yields the best plasmonic results, we next analyze the effect of the film thickness. Figure 3 shows the reflectance spectra for the three samples with a good agreement between the simulated and the experimental spectra, with plasma frequencies in the 4000 cm$^{-1}$ range. The thickness of the thin films does not seem to affect significantly the plasma frequency, even though there is a small variation that may be related to experimental reproducibility.

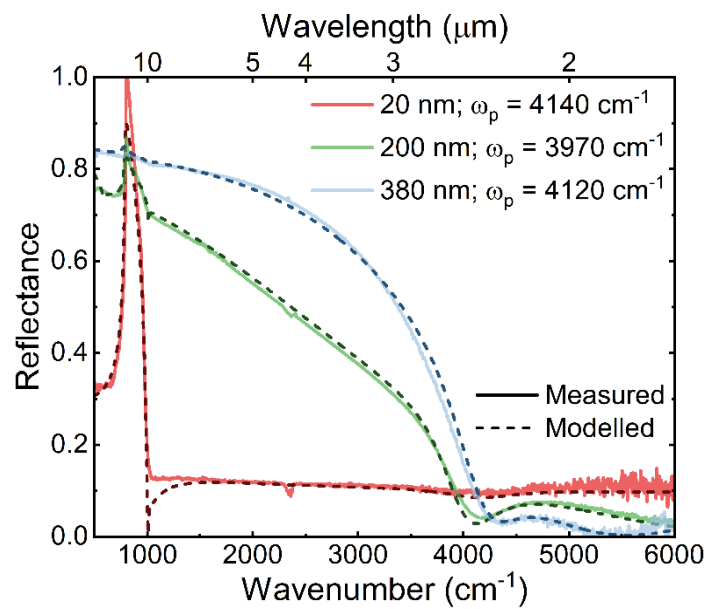

Figure 3: Reflectance of Set 2 of Cd(Zn)O samples with 10 % Zn, with three different Cd(Zn)O thicknesses. The solid line represents the experimental data, taken under p-polarization and an angle of incidence of 45°. The dashed line represents the fitted simulated data obtained with a TMM model. The legend shows the thickness and the plasma frequency obtained for each of the samples.

The formation of surface polaritons in the thin films was analyzed using attenuated total reflectance (ATR) in the Otto configuration (see Methods). As shown in Figure 4, we can observe two polariton modes in the experimental spectra, which are well reproduced by the TMM model using the material parameters previously extracted. The symmetric mode (also called long-range mode) is the result of the constructive coupling of the surface plasmon-polaritons at the interfaces of the system leading a large intensity of the electric field within the plasmonic film [29]. The antisymmetric mode (also called short-range mode) is characterized by the destructive coupling of the surface plasmon polaritons at the interfaces that result in a low electric field inside the plasmonic film and a significant magnification in the adjacent dielectrics. On a dielectric substrate without phonons, this mode's frequency varies between zero at small $k_x$ and $\omega_{spp,Sub}$ at large $k_x$, where $\omega_{spp,Sub}$ is the characteristic frequency of the surface plasmon polariton at the interface defined by equation (1).

$$\omega_{spp,sub} = \frac{\omega_p \cdot \sqrt{\varepsilon_{\infty,Cd(Zn)O}}}{\sqrt{\varepsilon_{\infty,Cd(Zn)O} + \varepsilon_{\infty,sub}}} \quad (1)$$

Here, $\varepsilon_{\infty,CdZnO}$ is the high frequency permittivity of Cd(Zn)O, and $\varepsilon_{\infty,sub}$ is the high frequency permittivity of the substrate. In contrast, on a polar substrate as SiC, the antisymmetric mode gets pinned at the LO phonon resonance at low $k_x$, and hybridizes into a highly propagative, low loss, surface plasmon-phonon polariton [23] whose frequency ranges between $\omega_{LO}$ and $\omega_{spp,SiC}$.

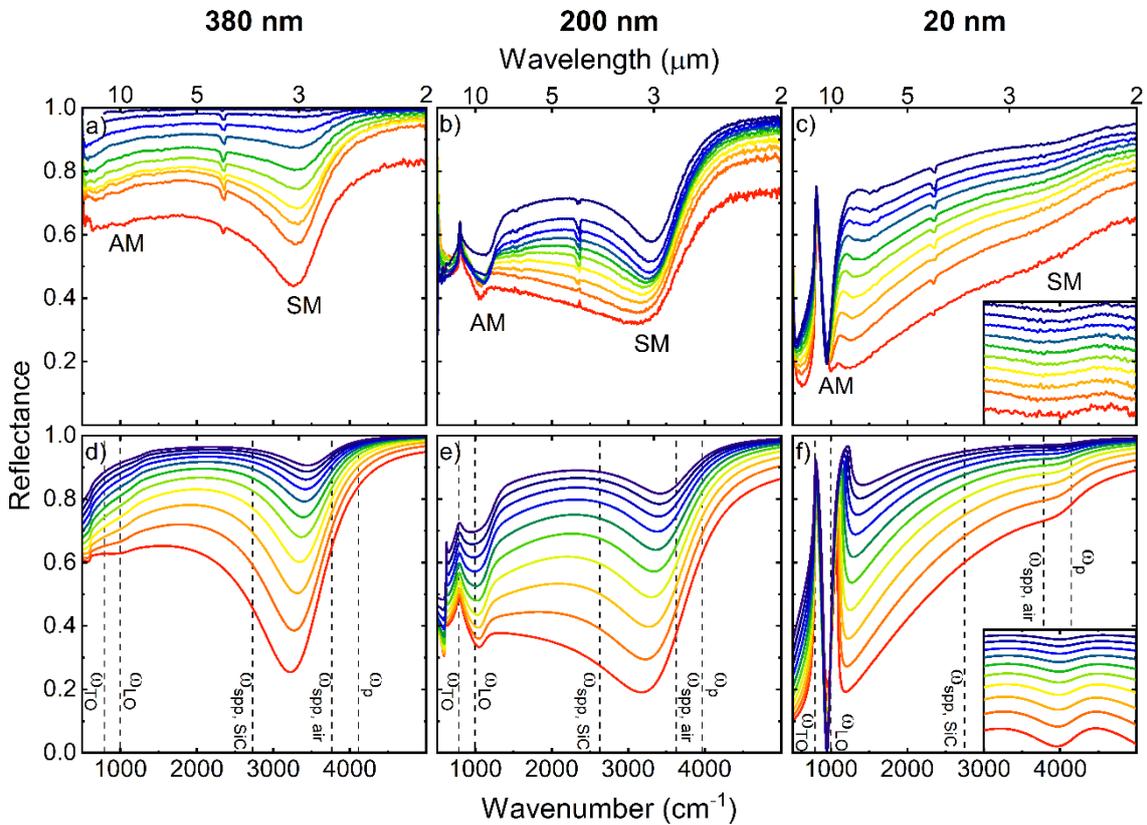

Figure 4: a-c) Experimental and d-f) simulated ATR spectra in p-polarization for the Cd(Zn)O samples with 10 % Zn and thickness of 20, 200 and 380 nm. The incident angle varies from 39 (light blue) to 57° (dark blue) in steps of 2°. The antisymmetric and symmetric modes are marked in each of the experimental reflectance spectra with the initials AM and SM respectively. Also plotted in the figure are the TO and LO phonon frequencies of SiC ($\omega_{TO,SiC}$ and $\omega_{LO,SiC}$, respectively), the surface plasmon polariton frequencies for the Cd(Zn)O/SiC surface and for the Cd(Zn)O/air surface ($\omega_{SPP,SiC}$ and $\omega_{SPP,air}$, respectively) and the Cd(Zn)O plasma frequency ($\omega_p$). The insets in figures c), f) show the minima around 3900 cm$^{-1}$ with a subtracted straight baseline to help visualizing them. The wavenumber in the insets spans from 3000 to 5000 cm$^{-1}$ in the same scale as their corresponding figures.

In order to analyze the frequency dispersion of the polariton modes, Figure 5 shows the simulated ATR contour plots for the Cd(Zn)O films as a function of in-plane momentum ($k_x$) for several film thicknesses. The mode frequencies extracted from the minima of the measured reflectance (Figure 4) are superimposed on the map as white dots. The figure clearly shows the two previously discussed modes. In the thickest sample the two modes are characterized by large dispersion. The symmetric mode tends towards the asymptote at $\omega_{spp,\,air}$ while the antisymmetric mode tends to the asymptote at $\omega_{spp,\,SiC}$ for large $k_x$. The decrease of the film thickness triggers a repulsion between the two modes, where the symmetric mode moves towards higher energies while the antisymmetric mode shifts to lower energies. Indeed, in the sample with 20 nm thickness, the strong interaction between the SPPs in each interface causes the two modes to separate drastically. The symmetric mode becomes nearly flat and gets pinned close to $\omega_p$. As seen in Figure S1a of the Supplemental Information, the mode falls inside the ENZ range in Cd$_{0.9}$Zn$_{0.1}$O (3700-4500 cm$^{-1}$). Meanwhile, in the same sample the hybridized antisymmetric mode experiences very low dispersion. As can be seen in the dispersion curves in Figure 5, the mode at low values of $k_x$ remains mainly flat and pinned at the LO frequency of SiC. Figure S1b shows how the mode only remains inside the SiC ENZ range at low values of $k_x$.

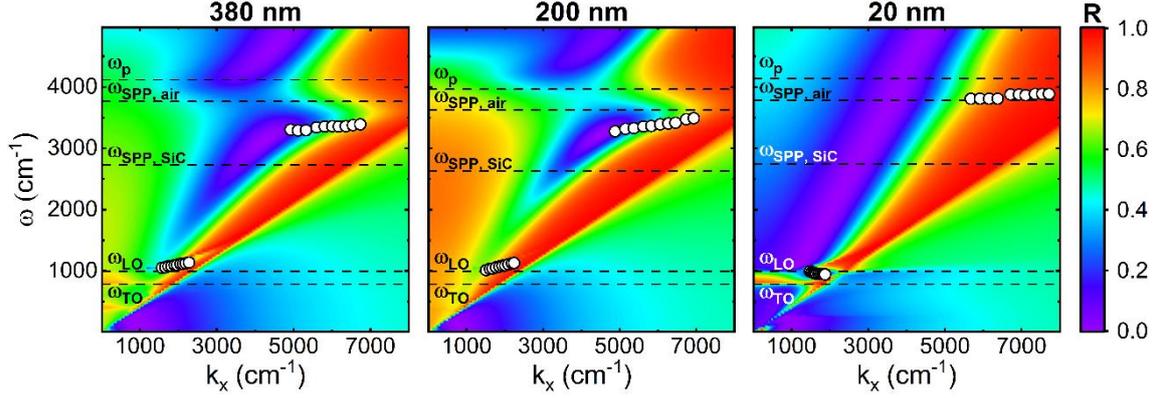

Figure 5: Simulated dispersion for the ATR contours of Cd(Zn)O films with 10 % Zn and thicknesses of 20, 200 and 380 nm. The white circles are the experimental ATR minima. The dashed black lines represent the SiC TO and LO phonon frequencies ($\omega_{TO}$ and $\omega_{LO}$, respectively), the surface plasmon polariton frequencies for the Cd(Zn)O/SiC surface and for the Cd(Zn)O/air surface ($\omega_{SPP,SiC}$ and $\omega_{SPP,air}$, respectively) and the Cd(Zn)O plasma frequency ($\omega_p$).

A property of ENZ polaritonic modes is that they show low group velocity [30]. We now verify if that is the case for both symmetric and antisymmetric modes. In order to extract this parameter, we have calculated their dispersion curves by evaluating the imaginary part of the Fresnel reflection coefficient $r_p$ with a TMM model. From these dispersion curves, the group velocity relative to the speed of light ($v_g/c = d\omega/dk_x$) can then be determined. Figure 6 shows the simulated dispersion curves for both modes and the extracted relative group velocities.

As seen in Figure 6a, the symmetric mode is observed at $\omega_{spp}$ for the thickest film (400 nm) and rises to $\omega_p$ as the thickness is decreased down to 20 nm. For all thicknesses, the group velocities are very close to zero for $k_x$ above 4000 cm$^{-1}$, and especially for our measurable range of $k_x$, as it is common for an ideal ENZ mode. A similar behavior is expected to happen when Cd(Zn)O is grown on other substrates, and particularly on sapphire.

However, in the case of the antisymmetric mode, the dispersion curves, and hence the group velocities, are heavily affected by the phononic character of the substrate. Figure 6b shows the calculated dispersion for the antisymmetric mode for both SiC and sapphire substrates, where the resonance ranges from $\omega_{LO}$ to the SPP frequency of the C(Zn)O/substrates interface ($\omega_{spp,sub}$) defined by equation (1). Considering that the high frequency permittivity is 5.1 [27], 6.5 [31] and 3.1 [32], for Cd(Zn)O, SiC and sapphire, respectively, we can calculate $\omega_{spp,sub}$ for each of the substrates. As a result of the much higher high-frequency permittivity in SiC, $\omega_{spp,sub}$ is 2670 cm$^{-1}$, much lower than in sapphire, where it reaches 3180 cm$^{-1}$. This causes the dispersion curve to flatten, yielding group velocities in SiC that, for our measurable values of $k_x$, are around half of what is obtained in sapphire (Figure 6b). For a 20 nm-thick Cd(Zn)O film over SiC, the relative group velocity is 0.04, and even if we increase the thickness to 400 nm the maximum relative group velocity is only 0.24. Hence, the antisymmetric mode in Cd(Zn)O/SiC has low relative group velocities, making it an excellent ENZ mode.

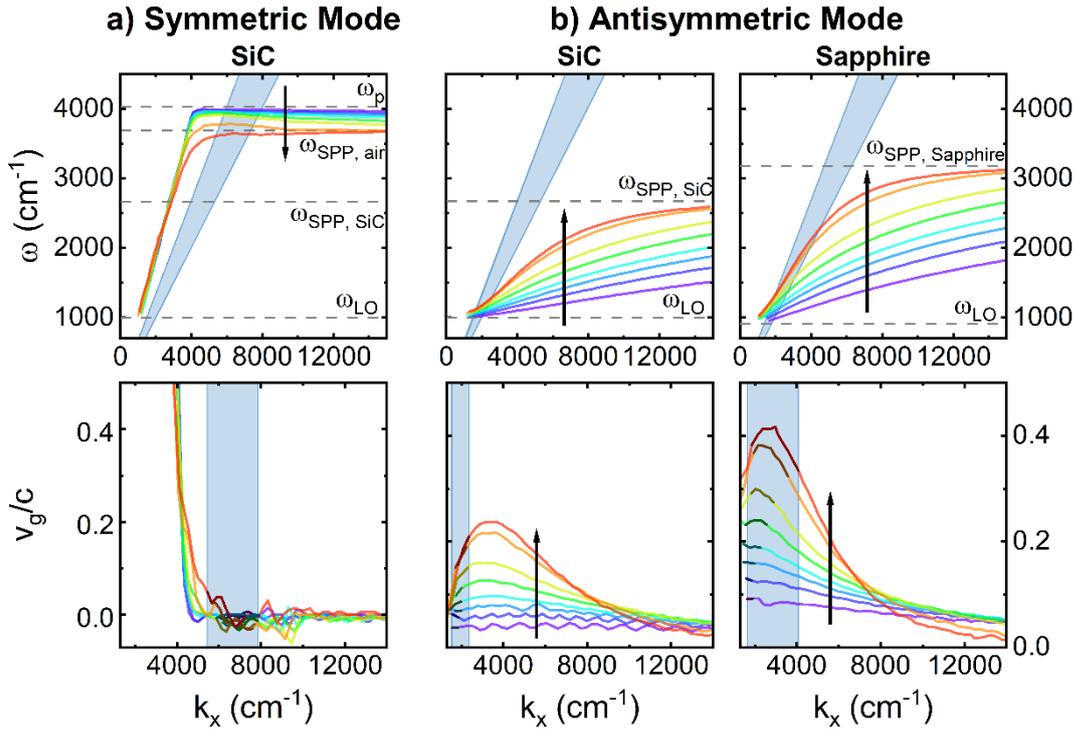

Figure 6: a) (top) Simulated dispersion curves of the symmetric mode for various Cd(Zn)O thicknesses from 20 (violet) to 400 (red) nm on a SiC substrate. The simulations are made with $\omega_p = 4030$ cm$^{-1}$ and $\gamma_p = 1$ cm$^{-1}$ for visibility. The blue cone represents our measurable region. a) (bottom) Group velocities from the dispersion curves of the symmetric mode in a SiC substrate. The dark curves represent our measurable range for each of the thicknesses, and the blue rectangle represents the whole range of our measurable $k_x$ b) (top) Simulated dispersion curves of the antisymmetric modes for thicknesses from 20 to 400 nm in SiC and sapphire substrates. b) (bottom) Group velocities from the dispersion curves of the antisymmetric mode in SiC and sapphire substrates. The dark curves represent our measurable range for each of the thicknesses, and the blue rectangle represents the whole range of our measurable $k_x$.

## 3. Conclusions

In this work we have shown that Cd(Zn)O thin films can be integrated on SiC yielding low-loss ENZ modes with excellent figures of merit in the 2.5 µm spectral region. The thin films have thicknesses ranging from 460 to 20 nm, plasma frequencies from 3100 to 4200 cm$^{-1}$ depending on the Zn content, and optical losses as low as 13 %. On these Cd(Zn)O/SiC structures two surface plasmon polariton resonances are observed that correspond to the symmetric and antisymmetric modes. The symmetric mode, a purely plasmonic resonance, is observed at a frequency that shifts to $\omega_p$ as the thickness is decreased, eventually pinning for the thinnest film with 20 nm, and showing in the ENZ range a relative group velocity extracted from the dispersion curves very close to zero. The antisymmetric mode is the result of the hybridization of the plasmonic mode from Cd(Zn)O and the high energy LO phonon from SiC, pinning at $\omega_{LO}$ for low $k_x$ values, and tending to $\omega_{spp,SiC}$ for large values. However, as a result of the high value of the high frequency permittivity in SiC, $\omega_{spp,SiC}$ is much lower than in sapphire, the widely used substrate for Cd(Zn)O, and the resulting relative group velocities for this mode are clearly lower, reaching values as low as 0.04 for the 20 nm-thick film. Taking together all these results, we show that SiC is an excellent candidate as a substrate for Cd(Zn)O, offering full access to plasmonic structures integrated with active electronic devices .

## 4. Methods

**Sample growth**

Cd(Zn)O (111) epilayers with nominal Zn concentrations ranging from 0 to ~20% were grown on c-plane (0001) 4H-SiC substrates. For reference and comparison, (001)-oriented Cd(Zn)O layers with similar nominal compositions were grown on r-plane sapphire (Al$_2$O$_3$) under comparable growth conditions. Dimethylcadmium (DMCd) and diethylzinc (DEZn) were used as cadmium and zinc sources, respectively, while tertiary butanol (TBA) served as the oxygen precursor. The growth temperature was maintained at 304°C in all experiments, and the precursor flow rates were adjusted to match nominal zinc compositions of 0%, 10%, and 20%, maintaining a TBA flow rate of 71.77 µmol/min and a II/VI precursors ratio approximatively (due to some limits in the precursors flow control) of 5.

**Morphological and Structural characterization**

Film morphology was characterized by scanning electron microscopy and atomic force microscopy. SEM top view and cross-section images were acquired using a Hitachi S-4800 microscope operating at 20 kV with a working distance of approximately 8 mm. AFM measurements were performed in dynamic mode with an AIST-NT SmartSPM system; samples were mounted on magnetic holders using carbon tape, without further surface preparation.

The chemical composition of the films was determined by energy-dispersive X-ray spectroscopy using the detector integrated in the Hitachi S-4800 microscope. An accelerating voltage of 20 kV was employed to ensure sufficient beam penetration and signal collection from the epilayers, as none of the relevant elements are present in the SiC substrate.

Structural characterization was carried out by high-resolution X-ray diffraction using a PANalytical X'Pert MRD diffractometer equipped with Cu Kα radiation ($\lambda$ = 1.54056 Å). Monochromatic, parallel-beam illumination was obtained using a parabolic mirror and a four-bounce hybrid monochromator in the incident beam path.

**Polarized reflectance spectroscopy measurements and ATR measurements**

Reflectance measurements were performed using a Fourier Transform Infrared (FTIR) spectrometer using p-polarized light with a 45° incident angle. The ATR technique in the Otto configuration was used with a ZnSe prism to access the surface polariton modes. The measurements in ATR were performed using p-polarized incident light ranging at angles of incidence between 39 and 59°.

**Numerical models**

A commercial software based on the TMM model was used to fit the reflectance in the samples and obtain the parameters of thickness, plasma frequency and damping. The dielectric functions of the two corresponding materials were modelled using the following equations:

$$\varepsilon_{Cd(Zn)O} = \varepsilon_{\infty, Cd(Zn)O} - \frac{\omega_p^2 \cdot \varepsilon_{\infty,Cd(Zn)O}}{\omega^2 + i\omega\gamma_{TO}}$$

$$\varepsilon_{SiC} = \varepsilon_\infty^{SiC} \frac{\omega_{LO}^2 - \omega^2 - i\omega\gamma_{LO}}{\omega_{TO}^2 - \omega^2 - i\omega\gamma_{TO}}$$

The dispersion curves were calculated using a homemade program based on the TMM model in p-polarization. The imaginary part of the reflection coefficient was obtained using the same program for models with varying thicknesses (20-400 nm), similar plasma frequency ($\omega_p = 4030$ cm$^{-1}$) and damping ($\gamma_p = 1$). The maxima in the resulting dispersion curves were extracted using the free software Webplotdigitizer.

## Declaration of competing interests

The authors declare that they have no competing interests that could influence the work reported in this paper.

## Acknowledgements

This work was partly supported by the following projects: PID2024-156706OB-C2, funded by MICIU/AEI/10.13039/ 501100011033/FEDER, UE; PDC2023-145827-C2 funded by MICIU/AEI/10.13039/501100011033 and by European Union Next Generation EU/ PRTR. The authors also acknowledge the support provided by the Servei Central de Suport a la Investigació Experimental (SCSIE), University of Valencia, for their instrumental facilities, and by the MICRONANOFABS ICTS network from MICIU.

## Data availability

The data that support the findings of this study are available from the corresponding author upon request.

# Supporting information of:

# Cd(Zn)O on SiC: epsilon-near-zero modes and plasmon-phonon coupling

Maria Villanueva-Blanco (maria.villanueva@upm.es) [a], Javier Yeste (Javier.Yeste@uv.es) [b], Julia Ingles-Cerrillo (julia.ingles@upm.es) [a], Pablo Ibañez-Romero (p.iromero@upm.es) [a], Carmen Martínez-Tomas (Carmen.Martinez-Tomas@uv.es) [b], Vicente Muñoz-Sanjosé (Vicente.Munoz@uv.es) [b], Miguel Montes Bajo (miguel.montes@upm.es) [a*] and Adrian Hierro (adrian.hierro@upm.es) [a*]

[a] ISOM, Universidad Politécnica de Madrid, 28040 Madrid, Spain

[b] Dept. Física Aplicada i Electromagnetisme, Universitat de València, Burjassot, Valencia, Spain

* To whom correspondence should be addressed.

## SUPPLEMENTARY

**S1. Permittivity and ENZ region in Cd(Zn)O and SiC**

The permittivities in Cd(Zn)O and SiC were calculated with the real losses obtained from the fitting the experimental reflectance with the model using equations (2) and (3), respectively. The real and imaginary parts of the dielectric functions are plotted in Figure S1 for both materials. The ENZ regions ($|Re(\varepsilon)| < 1$) are shaded in both figures. As seen in the Figure, the ENZ range in Cd(Zn)O spans from 3700-4500 cm$^{-1}$. At the zero-crossing, the imaginary part of the permittivity has a value $\text{Im}(\varepsilon) = 0.94$. On the other hand, the ENZ region in SiC spans from 970-1030 cm$^{-1}$ and the imaginary part of the permittivity has a value $Im(\varepsilon) = 0.23$ at the zero-crossing frequency, ensuring low losses.

The red and green circles in the figure represent the experimental minima obtained for the symmetric and antisymmetric modes, respectively, in the 20 nm thin film. As we can see, while the symmetric mode gets pinned in thin films in the ENZ mode, only the resonances for low $k_x$ fall inside the ENZ range in the antisymmetric mode.

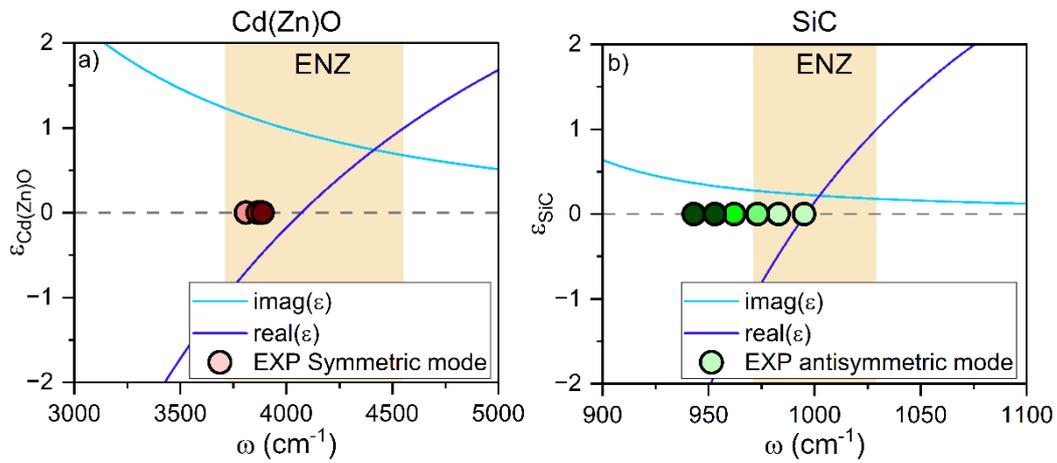

Figure S1: a) Permittivity calculated by the Drude model equations in Cd(Zn)O. The orange shaded region represents the ENZ range. The red circles represent the experimental minima for the symmetric mode in the 20 nm thin film where the lightest colors correspond to low $k_x$ and the darkest ones to high $k_x$. b) Permittivity calculated by the Gervais model equations in SiC. The orange shaded region represents the ENZ range. The green circles represent the experimental minima for the antisymmetric mode in the 20 nm thin film where the lightest colors correspond to low $k_x$ and the darkest ones to high $k_x$.